\documentclass[doublecol]{epl2}

\usepackage{amsmath}

\begin{document} 

\title{Reference data for phase diagrams of triangular and hexagonal 	
	bosonic lattices}

\shorttitle{Reference data for triangular and hexagonal bosonic lattices}

\author{N.~Teichmann\thanks{E-mail: 
	\email{Teichmann@theorie.physik.uni-oldenburg.de}} 
\and D.~Hinrichs 
\and M.~Holthaus}

\shortauthor{N.~Teichmann \etal}

\institute{Institut f\"ur Physik, 
	Carl von Ossietzky Universit\"at, 
        D-26111 Oldenburg, Germany}

\pacs{03.75.Lm}{Tunneling, Josephson effect, Bose-Einstein condensates 
	in periodic potentials}
\pacs{64.70.Tg}{Quantum phase transitions}
\pacs{67.85.Hj}{Bose-Einstein condensates in optical potentials}

\abstract{
We investigate systems of bosonic particles at zero temperature in triangular 
and hexagonal optical lattice potentials in the framework of the Bose-Hubbard
model. Employing the process-chain approach, we obtain accurate values for 
the boundaries between the Mott insulating phase and the superfluid phase.
These results can serve as reference data for both other approximation schemes
and upcoming experiments. Since arbitrary integer filling factors~$g$ are 
amenable to our technique, we are able to monitor the behavior of the critical 
hopping parameters with increasing filling. We also demonstrate that the 
$g$-dependence of these exact parameters is described almost perfectly by a 
scaling relation inferred from the mean-field approximation. 
}

\maketitle

\section{Introduction} 

Over the last ten years ultracold atoms in optical lattices induced by 
standing waves of laser radiation have become an outstandingly important 
and intensely studied testing ground for quantum many-body 
physics~\cite{Lewenstein2007,BlochDalibardZwerger2008}. Great prospects 
offered by these systems stem from the chance to investigate condensed-matter 
phenomena by simulating paradigmatic model Hamiltonians in the 
laboratory~\cite{BulutaNori2009}. In particular, the Bose-Hubbard 
Hamiltonian~\cite{Fisher1989,Jaksch1998} has attracted a lot of attention, 
since it describes ultracold bosonic atoms in an optical lattice potential 
fairly well. This system exhibits a quantum phase transition from a superfluid 
to a Mott insulator upon increasing the lattice depth~\cite{Greiner2002,
Zwerger2003}. Its extensions even show further interesting phases, 
{\em e.g.\/} a supersolid state~\cite{Leggett1970}, when admitting 
particle-particle interactions between neighboring sites~\cite{Scarola2006} 
or introducing Bose-Fermi mixtures~\cite{Titvinidze2008}.

So far, most studies dealing with the Bose-Hubbard model have considered 
a square or a cubic lattice. For these particular lattice geometries the 
superfluid-insulator phase boundary has been calculated by various 
methods, such as mean-field approaches~\cite{Fisher1989,Kampf1993,
Bruder1993,vanOosten2001,vanOosten2003,Schroll2004}, the quantum rotor 
approach~\cite{Polak2007}, or a variational cluster 
formulation~\cite{Knap2010}. Arguably, the most precise results have been 
achieved by the strong coupling expansion~\cite{FreericksMonien1996,
ElstnerMonien1999,FreericksEtAl2009} and by Quantum Monte Carlo 
simulations~\cite{Capogrosso2007,Capogrosso2008} for low filling factors 
of the lattice, and by means of the process-chain approach for arbitrarily 
high integer filling~\cite{TeichmannEtAl2009_1}.
 
Quite recently, the successful experimental realization of planar triangular 
and hexagonal lattices has been reported~\cite{BeckerEtAl2009}. However, 
reliable theoretical data for the phase boundaries pertaining to these 
lattice types still seem to be missing, except for the single case of a 
triangular lattice at unit filling  ($g=1$), which has been covered by a 
strong coupling expansion~\cite{ElstnerMonien_arxiv1999}. Apart from the 
need to compare experimental results to accurate theoretical predictions,
precise knowledge of the critical values of the hopping parameters would
also be of great value to aid the development of new approximation schemes,
and of future numerical methods.  

\begin{figure}
\centering
\includegraphics[scale=0.4]{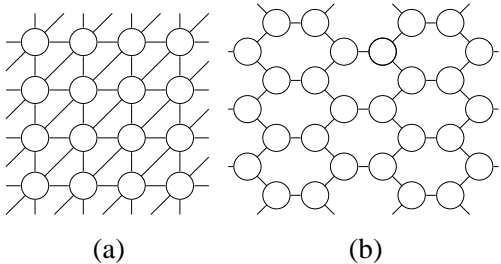}
\caption{Schematic illustration of the triangular~(a) and the hexagonal~(b) 
	lattice. Circles represent lattice sites occupied by bosonic 
	particles; lines represent nearest-neighbor couplings due to 
	tunneling processes between adjacent sites.}
\label{fig:lattices}
\end{figure}

In this contribution we provide the phase diagrams for the Bose-Hubbard model 
with planar triangular and hexagonal lattice geometries. These two lattice 
types are depicted schematically in fig.~\ref{fig:lattices}. The process-chain
approach~\cite{Eckardt2009} in combination with the method of the effective
potential~\cite{NegeleOrland1998,dosSantos2009} enables us to compute the
phase boundaries with high precision, as has been demonstrated previously
for square and cubic lattices~\cite{TeichmannEtAl2009_1,TeichmannEtAl2009_2}.

For self-consistency, we start with a brief description of the Bose-Hubbard 
model, and give a short explanation of both the process-chain approach and 
the method of the effective potential, which provides the signature of the 
phase transition. We then present our results for the phase diagrams arising 
from triangular and hexagonal lattices, and state the corresponding critical 
values of the hopping parameter $(J/U)_{\rm c}$ and of the chemical potential 
$(\mu/U)_{\rm c}$. Since we can treat lattices with an arbitrary number of
particles per lattice site, {\em i.e.\/,} with an arbitrarily high integer
filling factor~$g$, we are able to reveal that the critical values 
$(J/U)_{\rm c}$ can be scaled such that they become (almost) independent 
of the filling factor.

\section{The model} 

We study the homogeneous Bose-Hubbard model, given by the Hamiltonian
\begin{equation}
	H = \frac{U}{2} \sum_{i}
	\hat{n}_i(\hat{n}_i-1) 
	- J \sum_{\langle i,j \rangle} \, \hat{a}_i^{\dagger} 
	\hat{a}_j^{\phantom \dagger}
	-  \mu \sum_{i} \hat{n}_i
        \;,
\label{eq:Hamiltonian_full}
\end{equation} 
which embodies in an elementary way the competition between the kinetic 
energy due to tunneling processes and the potential energy associated with
the repulsive interaction of bosons on the same lattice site. The operators 
$\hat{a}_i^{\phantom \dagger}$ and $\hat{a}_i^{\dagger}$ are the bosonic 
annihilation and creation operators at site No.~$i$, and $\hat{n}_i$ is the 
corresponding number operator. We examine the case of zero temperature, 
which permits a single-band description, such that one only needs to consider 
the lowest Wannier state at each site. Moreover, an on-site approximation is 
made here, assuming that only particles sitting on the same lattice site 
interact with each other, each on-site pair contributing the amount~$U$ to
the total interaction energy. Hopping processes of the bosons are restricted 
to adjacent sites; their strength is quantified by the matrix element~$J$. 
The subscript ${\langle i,j \rangle}$ at the kinetic-energy sum indicates 
that this summation only includes pairs of neighboring sites. For the 
homogeneous systems studied here, the chemical potential~$\mu$ is constant 
throughout the lattice.

When expressing all energies in multiples of the on-site pair interaction
energy ~$U$, we arrive at the dimensionless Hamiltonian 
\begin{equation}
	H_{\rm BH} = \frac{1}{2} \sum_{i}
	\hat{n}_i(\hat{n}_i-1) 
	- J/U \sum_{\langle i,j \rangle} \, \hat{a}_i^{\dagger} 
	\hat{a}_j^{\phantom \dagger}
	-  \mu/U \sum_{i} \hat{n}_i	
\label{eq:Hamiltonian_dimensionless}
\end{equation} 
containing two parameters, the hopping parameter $J/U$ and the scaled 
chemical potential $\mu/U$.

The existence of a quantum phase transition from a Mott insulator to a 
superfluid~\cite{Fisher1989,Greiner2002} in response to an increase of the 
hopping parameter is made plausible by inspecting the limiting cases: When 
\mbox{$J/U \gg 1$} one has an almost ideal Bose-Einstein condensate with all 
particles occupying the zero-quasimomentum Bloch state. This corresponds to 
a superfluid with all particles delocalized, and phase fluctuations being 
suppressed. The superfluid phase is characterized by long-range order and 
non-zero compressibility, 
$\partial\left\langle n\right\rangle/\partial\mu \neq 0$. 
In the opposite limit $J/U \ll 1$ hopping is prohibited and all sites are 
decoupled from each other, so that the 
Hamiltonian~(\ref{eq:Hamiltonian_dimensionless}) becomes diagonal in the 
occupation number basis. Minimizing the on-site energy, one finds that an
integer number $g = N/M$ occupies each site, with $N$ denoting the total 
number of particles, and $M$ the number of lattice sites. This phase is 
characterized by reduced density fluctuations and incompressibility, 
{\em i.e.\/} $\partial\left\langle n\right\rangle/\partial\mu = 0$. 
The ground state $| m \rangle$ for $J/U=0$ simply is a product state of Fock 
states with $g$~particles on each site,
\begin{equation}
	| m \rangle = \prod_{i=1}^M \frac{\left(\hat{a}_i^{\dagger}\right)^g}
        {\sqrt{g!}} |0 \rangle \; ,
\label{eq:GroundStateMI} 
\end{equation} 
where $|0 \rangle$ is the particle-free vacuum. When starting in the 
Mott-insulating phase and increasing $J/U$ from zero to higher values 
for a given, fixed chemical potential $\mu/U$, there is a value 
$(J/U)_{\rm pb}$ at which the excitation gap vanishes, marking the 
entrance into the superfluid regime.

\section{The Method} 

In order to determine these values $(J/U)_{\rm pb}$ of the hopping parameter 
at the phase boundary we make use of the method of the effective
potential~\cite{NegeleOrland1998,dosSantos2009,TeichmannEtAl2009_2}, which 
requires to add source and drain terms of constant strength $\eta$ and 
$\eta^*$ to the Bose-Hubbard Hamiltonian~(\ref{eq:Hamiltonian_dimensionless}):
\begin{equation}
	\tilde H_{\rm BH}(\eta, \eta^*) = H_{\rm BH} 
	+ \sum_i  \left(\eta^* \hat{a}_i^{\phantom \dagger} 
	+ \eta \hat{a}_i^{\dagger}\right) \; .
\label{eq:Hamiltonian_sourcedrain}
\end{equation}
The corresponding grand canonical free energy
\begin{equation}
	F(J/U,\eta, \eta^*) = 
	M \left(f_0(J/U) + \sum_{n=1}^{\infty} c_{2n}(J/U) |\eta|^{2n} \right)	
\label{eq:free_energy}
\end{equation}
with expansion coefficients 
\begin{equation}
	c_{2n}(J/U) = \sum_{\nu=0}^{\infty} \alpha_{2n}^{(\nu)} (J/U)^{\nu}	
\label{eq:series_c2}
\end{equation}
then is Legendre-transformed into an effective potential
\begin{equation}
	\Gamma( J/U, \psi, \psi^*) = F_0 - \frac{1}{c_2}|\psi|^2 
	+\frac{c_4}{c_2^4}|\psi|^4 + \mathcal{O}(|\psi|^6) \; .
\label{eq:effpot}
\end{equation} 
Odd orders of $\eta$ vanish in the expansion~(\ref{eq:free_energy}) 
of the free energy because of the $U(1)$-symmetry of the augmented 
Hamiltonian~(\ref{eq:Hamiltonian_sourcedrain}). The expansion parameter 
$|\psi|^2$ of the effective potential~(\ref{eq:effpot}) serves as the
order parameter; it is given by
\begin{eqnarray}
	\psi(\eta) &= \displaystyle{\frac{1}{M}
	\frac{\partial F} {\partial \eta^*}}
	&= \langle \hat{a}_i^{\phantom \dagger} \rangle_{\eta}\;,
	\nonumber\\
	\psi^*(\eta) &= \displaystyle{\frac{1}{M}
	\frac{\partial F}{\partial \eta}}
	&= \langle \hat{a}_i^{\dagger} \rangle_{\eta} \; .
\label{eq:order_parms}
\end{eqnarray} 
The Legendre pair $\eta$ and $\psi^*$ obeys the identity  
\begin{equation}
	\frac{\partial\Gamma}{\partial \psi^*} = -\eta \; ;
\label{eq:Gamma_psi}	
\end{equation}
the complex conjugate of this equation connects $\eta^*$ and $\psi$. Now the
original Bose-Hubbard Hamiltonian~(\ref{eq:Hamiltonian_dimensionless}) is
recovered from the augmented Hamiltonian~(\ref{eq:Hamiltonian_sourcedrain})
by setting $\eta = \eta^* = 0$, which means that the relevant values of
$\psi$ and $\psi^*$ are those which render the effective potential $\Gamma$ 
stationary. For low hopping strengths $J/U$, when the system is in its
Mott-insulating phase, the coefficient $c_2$ in the 
expansions~(\ref{eq:series_c2}) and (\ref{eq:effpot}) is negative, whereas 
$c_4$ is posi\-tive, leading to a minimum of $\Gamma$ at $|\psi|^2 = 0$.
The order parameter $|\psi|^2$ adopts a non-zero value in the superfluid 
phase, signaling long-range order. The phase transition therefore takes  
place at that value of $J/U$ for which $1/c_2$ vanishes, so that the minimum 
of the expression~(\ref{eq:effpot}) starts to deviate from $|\psi|^2=0$. 
The upshot is that the phase boundary $(J/U)_{\rm pb}$ equals the radius of 
convergence of the series~(\ref{eq:series_c2}) for the coefficient $c_2$.

The coefficients $\alpha_2^{(\nu)}$ of that series are calculated within 
the process-chain approach, which is based on a diagrammatic
evaluation~\cite{TeichmannEtAl2009_2,Eckardt2009} of Kato's perturbation 
series~\cite{Kato1949}. The Kato formula for the $n$th-order energy 
correction experienced by a nondegenerate unperturbed state $|m\rangle$ in 
response to a perturbation~$V$ reads 
\begin{eqnarray}	
	E_m^{(n)}
	= {\rm tr} \left[ \sum_{ \{\alpha_\ell\} } 
	S^{\alpha_1} V S^{\alpha_2} V S^{\alpha_3} \ldots 
	S^{\alpha_n}VS^{\alpha_{n+1}}  \right] \; .
\label{eq:Kato_eigenvalue}
\end{eqnarray}
Here the sum runs over all sets of $n+1$ non-negative integers $\alpha_\ell$ 
which obey the constraint $\sum_\ell \alpha_\ell = n-1$. The linking 
operators $S^{\alpha}$ are defined by
\begin{equation}
S^{\alpha} = \left\{\begin{matrix} 
	-| m\rangle \langle m| 
        &\quad\text{for } \alpha = 0 \\
	\displaystyle
	\sum\limits_{i\neq m} 
	\frac{|i \rangle \langle i|}{(E_m^{(0)} - E_i^{(0)})^{\alpha}} 
        &\quad\text{for } \alpha > 0 
	\end{matrix}\right. \; ,
	\label{eq:Kato_S}
\end{equation}
where $|i\rangle$ denotes the unperturbed ``intermediate'' eigenstates, 
and $E_i^{(0)}$ the corresponding unperturbed eigenvalues. Kato's trace 
formula~(\ref{eq:Kato_eigenvalue}) can be rewritten as a sum of matrix 
elements of the state $|m\rangle$ considered,
\begin{equation}        
        \langle m| V S^{\alpha_1} V S^{\alpha_2} \ldots 
        S^{\alpha_{n-1}} V |m \rangle \; .
\label{eq:Kato_term}
\end{equation}
The number of such matrix elements (\emph{Kato-terms}) quickly increases with 
the order~$n$ of perturbation theory. In first order, the only Kato-term is 
$\langle m| V |m\rangle$, while $n=2$ leads to $\langle m| V S^1 V |m\rangle$.
These are precisely the well known first- and second- order energy corrections,
as becomes obvious when inserting $S^1$ from eq.~(\ref{eq:Kato_S}). Each 
Kato-term~(\ref{eq:Kato_term}) can be viewed as a (sum of) closed process 
chain(s) consisting of $n$ processes caused by the perturbation~$V$, leading 
from the state $|m\rangle$ over various intermediate states $|i\rangle$ back 
to $|m\rangle$ again. When dealing with a homogeneous lattice system, many 
process chains can be combined into diagrams by appending an appropriate 
weight factor. This procedure drastically reduces the numerical effort. A more 
detailed description of the application of this process-chain technique to the 
Bose-Hubbard model is given in ref.~\cite{TeichmannEtAl2009_2}.

In our case, the unperturbed part of the Hamiltonian is site-diagonal,
reading
\begin{equation}
	H_0 = \frac{1}{2} \sum_{i}\hat{n}_i(\hat{n}_i-1) 
        - \mu/U \sum_{i} \hat{n}_i \; .
\label{eq:Hamiltonian_unperturbed}
\end{equation} 
The perturbation is given by the tunneling operators in combination with 
the source and drain terms artificially introduced in 
eq.~(\ref{eq:Hamiltonian_sourcedrain}): 
\begin{equation}
	V = - J/U \sum_{\langle i,j \rangle} \, \hat{a}_i^{\dagger} 
	\hat{a}_j^{\phantom \dagger}
	+ \sum_i  \left(\eta^* \hat{a}_i^{\phantom \dagger} 
	+ \eta \hat{a}_i^{\dagger}\right) \; .
\label{eq:Hamiltonian_perturbed}
\end{equation} 
Instead of using Kato's formulation for computing the total energy 
corrections, we employ it for calculating the coefficients $\alpha_2^{(\nu)}$ 
of the series~(\ref{eq:series_c2}) for $c_2$ only; the searched-for phase 
boundary $(J/U)_{\rm pb}$ then is determined in a second step as the radius 
of convergence of this series. Because $c_2$ is the coefficient of $|\eta|^2$ 
in the expansion of the free energy~(\ref{eq:free_energy}), it is associated 
with exactly one creation and one annihilation event of a particle. Hence, 
for calculating its coefficients $\alpha_2^{(\nu)}$ one has to evaluate 
only diagrams containing one creation (symbolized by a dot: $\bullet$) and 
one annihilation process ($\times$), together with $\nu$ tunneling processes 
($\rightarrow$).

\begin{figure}
\centering
\includegraphics[width=0.4\textwidth]{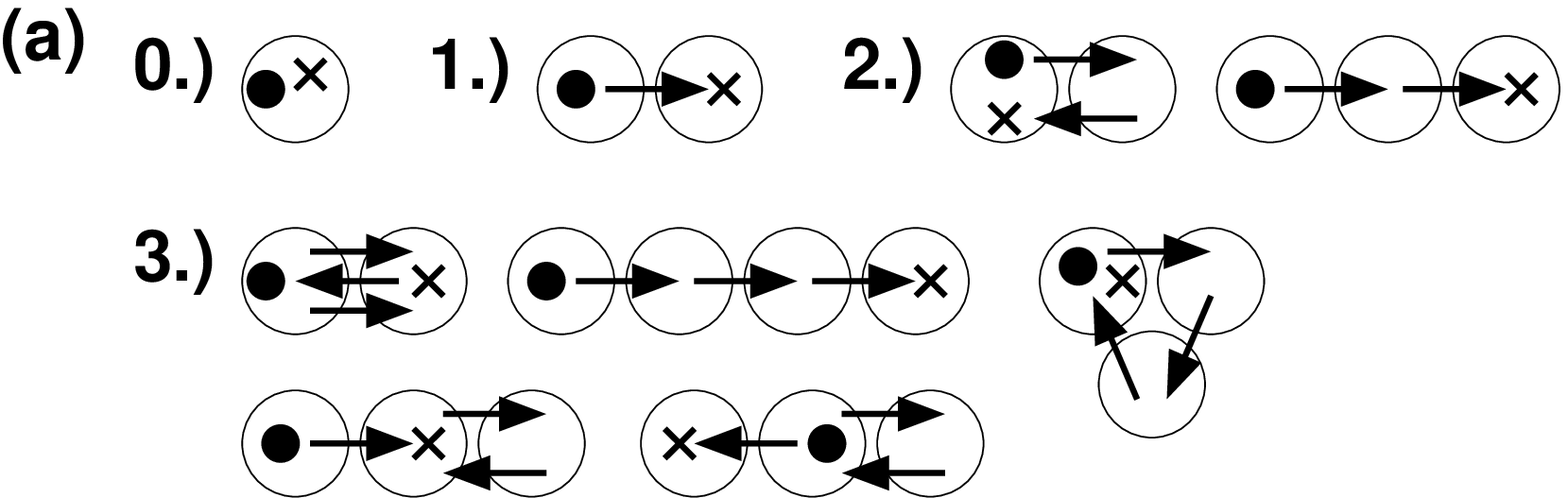}\\
\vspace{3mm}
\includegraphics[width=0.4\textwidth]{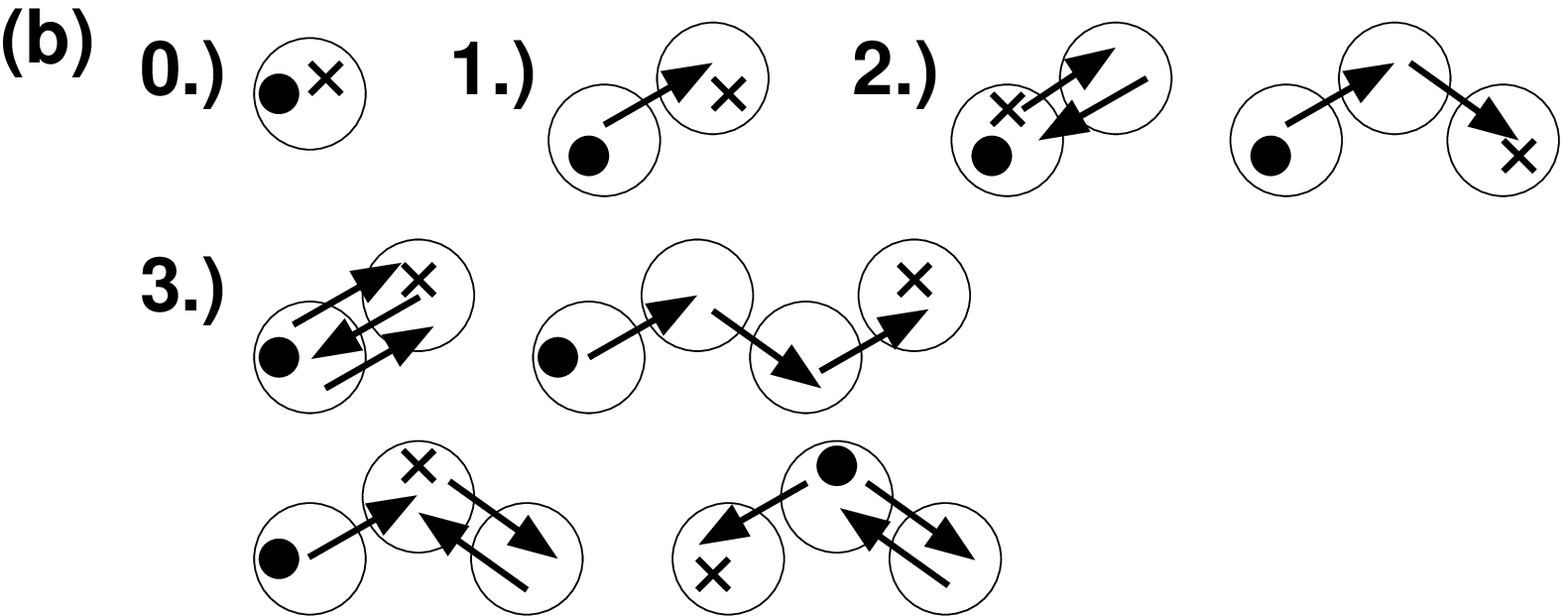}
\caption{Diagrams of order $\nu = 0, \ldots, 3$ in the hopping para\-meter $J/U$ 
	for a triangular and for a hexagonal lattice. The symbol~$\bullet$ 
	indicates the creation of a particle, $\rightarrow$ corresponds to 
	tunneling between two adjacent sites, and $\times$ to an annihilation 
	process. Upper figure (a): Diagrams for the triangular lattice; 
	weight factors are 0.) 1; 1.) 6; 2.) 6 and 30; 3.) 6, 138, 12, 30 
	and 30. Lower figure (b): Diagrams for the hexagonal lattice; 
	weight factors are 0.) 1; 1.) 3; 2.) 3 and 6;  3.) 3, 12, 6 and 6.}
 \label{fig:tria_hex_dias}
\end{figure}

Both the precise structure of the diagrams and their weight factors are 
determined by the geometry of the underlying lattice. 
Figure~\ref{fig:tria_hex_dias} lists the diagrams of order $\nu = 0, \ldots, 3$  
in the hopping parameter $J/U$ for a triangular and for a hexagonal lattice,
together with their respective weight factors. Each ``hexagonal'' diagram 
shown in (b) is topologically equivalent to a ``triangular'' one in (a),
but when taking three hopping processes into account a circular diagram
turns up in the triangular case which has no hexagonal counterpart. In 
higher orders of the hopping parameter the number of ``triangular'' diagrams 
even becomes much larger than that of the ``hexagonal'' ones,  as 
table~\ref{tab:p_tria_hexa} documents: The increase of the number of 
diagrams with the number $\nu$ of tunneling processes is much more 
pronounced in the triangular case. As another consequence of the geometric 
variation, the weight factors of corresponding diagrams generally differ
for the two lattice types.

\begin{table}
\centering
\begin{tabular}{c*{11}{|@{\hspace{3pt}}r@{\hspace{3pt}}}}
\hline 
$\nu $     & 0 & 1 & 2 & 3 &  4 &  5 &   6 &   7 &    8 &  9 & 10  \\ \hline 
Triangular & 1 & 1 & 2 & 5 & 14 & 41 & 129 & 416 & 1398 &  ×  &   ×  \\ 
Hexagonal  & 1 & 1 & 2 & 4 &  9 & 18 &  39 &  80 &  180 & 389 & 1260 \\
\hline
\end{tabular}
\caption{Number of diagrams to be evaluated when calculating the phase 
	boundary of the Bose-Hubbard model for a triangular and for a 
	hexagonal lattice to $\nu$th order in the hopping parameter~$J/U$, 
	corresponding to the order $\nu + 2$ of Kato's perturbation series.}
\label{tab:p_tria_hexa} 
\end{table}

The numerical value of a diagram is determined by going through all 
permutations of its individual constituent processes; for each permutation
one has to evaluate those Kato-terms which match it. The outcome then is
multiplied by the weight factor of the diagram in question. Finally the 
contributions of all diagrams occurring in a given order of perturbation 
theory are summed to yield the desired quantity $a_2^{(\nu)}$. For example, 
when considering the hexagonal lattice with $\nu = 3$ tunneling processes, 
four diagrams depicted in fig.~\ref{fig:tria_hex_dias}~(b) have to be 
dealt with. Each one of these leads to up to $5!=120$ different sequences 
of processes which have to be matched with 3 Kato-terms. Evidently the 
computational effort increases rapidly with the number~$\nu$ of tunneling 
processes taken into account: Both the number of Kato-terms and the number 
of diagrams proliferates quickly; in addition, the number of process 
permutations grows factorially with the order $n=\nu+2$ of perturbation 
theory.

\section{Results} 

\begin{figure}
\centering
\includegraphics[scale=0.32,angle=-90]{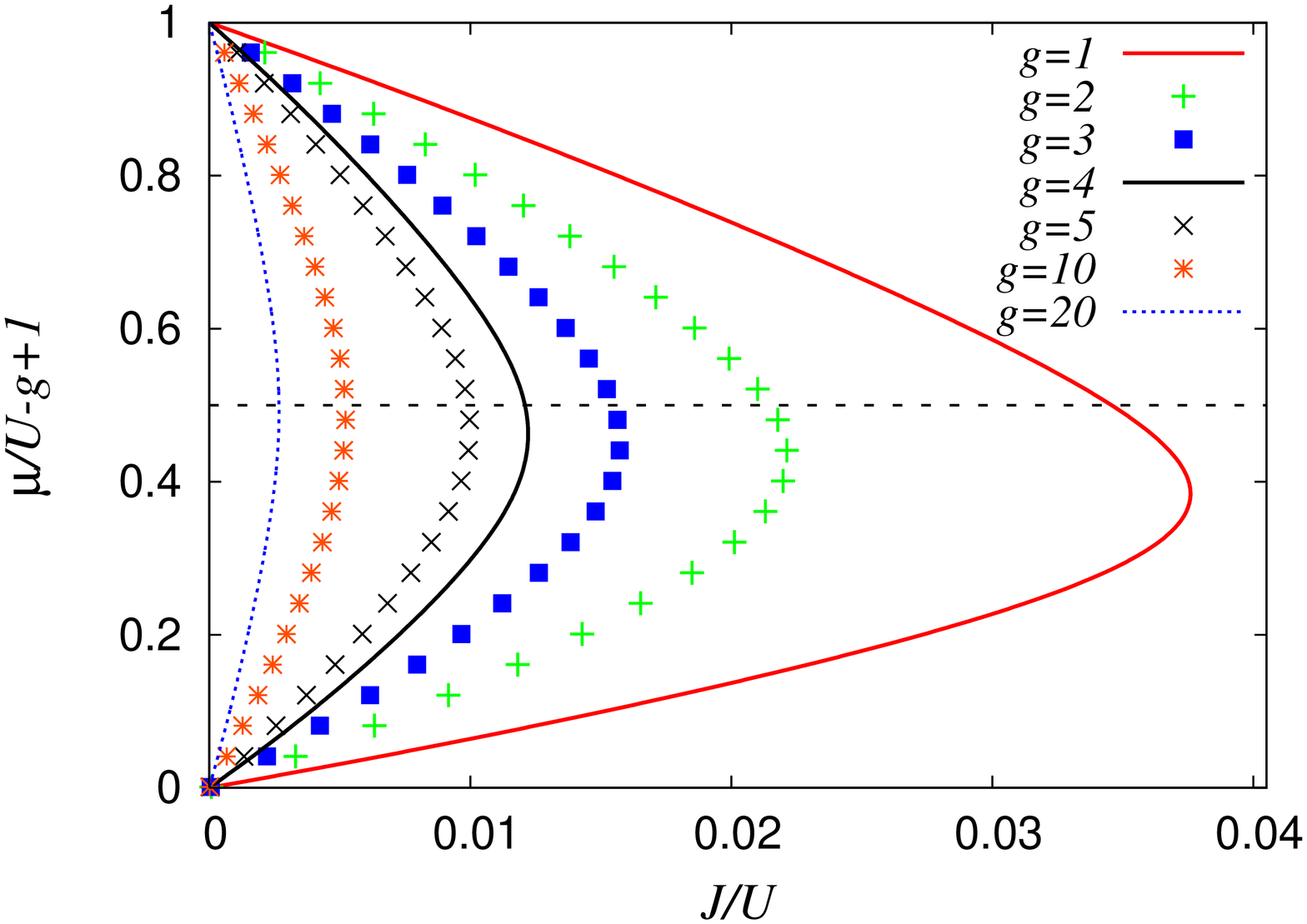}
\includegraphics[scale=0.32,angle=-90]{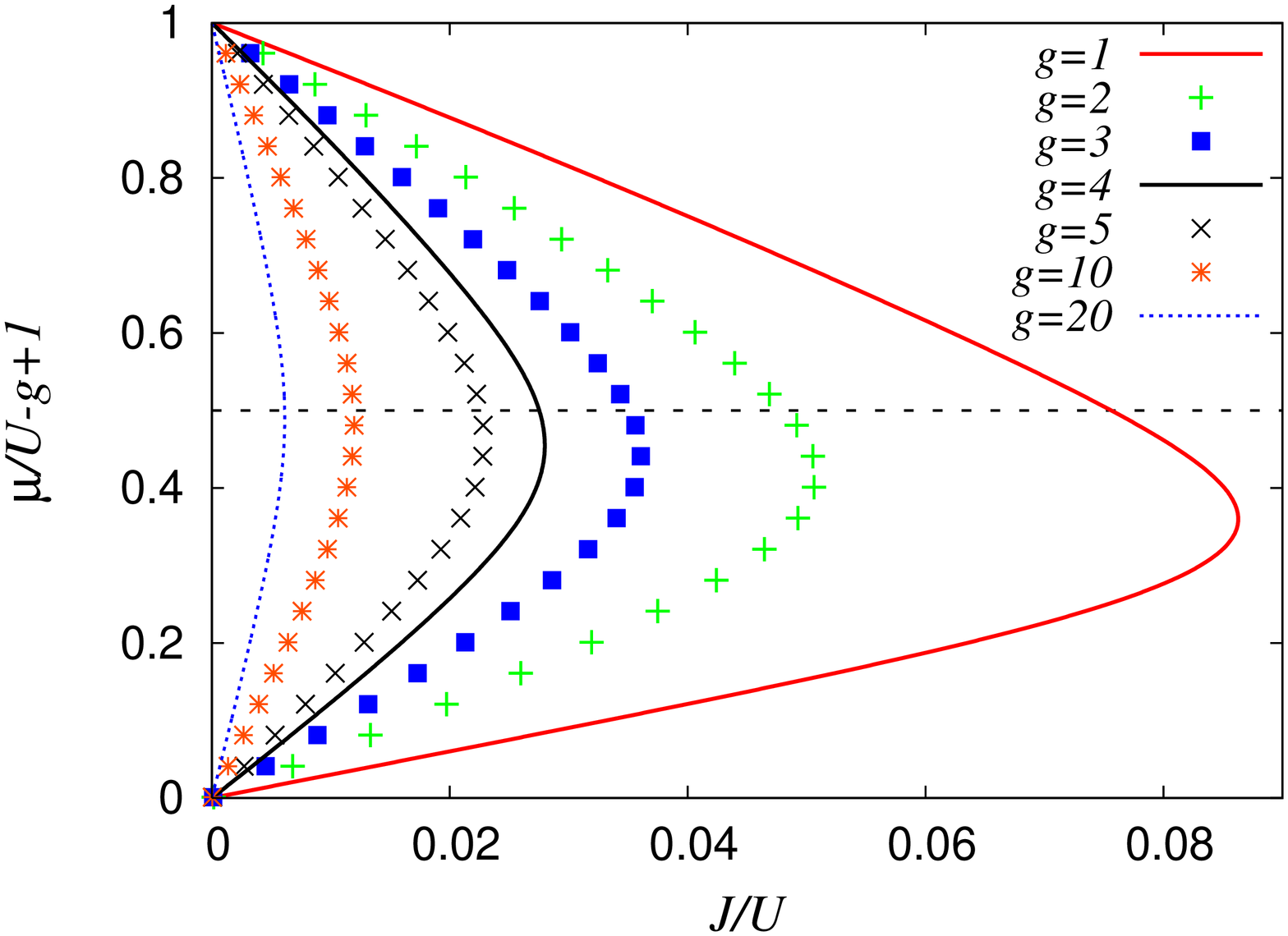}
\caption{(Color online) Mott lobes for the triangular lattice (upper panel) 
and for the hexagonal lattice (below) for various filling factors~$g$. 
The dashed horizontal line $\mu/U = g - 0.5$ marks the axis of particle-hole
symmetry which appears in the limit of large $g$.}
\label{fig:phasediagrams}
\end{figure}

For each preselected value of the chemical potential $\mu/U$, the corresponding 
coefficients $\alpha_2^{(\nu)}$ of the series~(\ref{eq:series_c2}) for $c_2$ 
show an almost geometric behavior, for both the triangular and the hexagonal 
lattice. As outlined above, the boundary $(J/U)_{\rm pb}$ between the Mott 
insulating and the superfluid phase is given by the lowest $J/U$ for which 
this series diverges. Thus, for delineating the phase boundary 
we determine its radius of convergence by means of d'Alembert's ratio 
test~\cite{WhittakerWatson}:
\begin{equation}
        (J/U)_{\rm pb} = \lim_{\nu\to \infty}\left|  
        \frac{\alpha_2^{(\nu-1)}}{\alpha_2^{(\nu)}}\right|
        \;.
\end{equation} 
The required extrapolation $\nu\to\infty$ is carried out by a linear fit of 
the ratios $\alpha_2^{(\nu-1)}/ \alpha_2^{(\nu)}$ over $1/\nu$; the desired 
value $(J/U)_{\rm pb}$ then is the point of intersection with the ordinate. 
This procedure also gives access to the relative error of $(J/U)_{\rm pb}$: 
Varying the set of coefficients $\alpha_2^{(\nu)}$ employed for the fit 
({\em e.g.\/}, taking only $\nu=4, \ldots, 8$) yields slightly different 
results; such fluctuations quantify the uncertainty of the final data.  
Here we employ the coefficients $\nu=2, \ldots, 8$, leading to an estimated 
relative error of less than $1$\%  for the triangular case, and about $2$\% 
for the hexagonal one. 

The phase diagrams for the two lattice types are plotted in  
fig.~\ref{fig:phasediagrams} in the $\mu/U$ vs.\ $J/U$-plane, for various 
filling factors~$g$. The critical values $(\mu/U)_{\rm c}$ and $(J/U)_{\rm c}$,
{\em i.e.\/} the chemical potential and the hopping parameter at the tip of 
the respective Mott lobe, are listed in table~\ref{tab:J_c}. Our result for 
the triangular lattice with unit filling compares favorably to the previous
finding of Elstner and Monien~\cite{ElstnerMonien_arxiv1999}: These authors 
have stated $(J/U)_{\rm c} = 0.037785$, whereas we obtain
$(J/U)_{\rm c} = 0.03759$; the deviation of about $0.5$\% is well within 
the estimated error margin. Because the coordination number $z_{\rm t} = 6$ 
of the triangular lattice is twice as large as that for the hexagonal one, 
$z_{\rm h} = 3$, the ``triangular'' critical hopping strength at unit filling 
is substantially lower --- by a factor of about $2.3$ --- than the 
``hexagonal'' one. On the other hand, despite the fact that the coordination 
number of the triangular lattice coincides with that of the simple 
three-dimensional (3D) cubic lattice, the corresponding critical hopping
strengths differ appreciably: The cubic lattice yields 
$(J/U)_{\rm c}\approx 0.0341$ for $g=1$, 
see refs.~\cite{Capogrosso2008,TeichmannEtAl2009_1}, amounting to a deviation
of approximately $9$\% from the triangular-lattice value. Inspecting the
Mott lobes in fig.~\ref{fig:phasediagrams}, one also confirms that the 
critical chemical potential $(\mu/U)_{\rm c}$ tends to $g-0.5$ with 
increasing filling factor~$g$, as expected from the particle-hole symmetry
which emerges in the large-$g$-limit.

\begin{table}
\begin{tabular}{r|r|r|r|r}
	\hline
      & \multicolumn{2}{c|}{Triangular} & \multicolumn{2}{c}{Hexagonal} \\ 
      	\hline
      	$g$ & $(\mu/U)_{\rm c}$ & $(J/U)_{\rm c}$ & $(\mu/U)_{\rm c}$ 
      & $(J/U)_{\rm c}$ \\ 
     	\hline
        1 & 0.384        & 3.759E-02   &  0.360       & 8.628E-02       \\
        2 & 1.432        & 2.214E-02   &  1.418       & 5.075E-02       \\
        3 & 2.452        & 1.574E-02   &  2.442       & 3.606E-02       \\
        4 & 3.463        & 1.222E-02   &  3.455       & 2.799E-02       \\
        5 & 4.469        & 9.984E-03   &  4.463       & 2.288E-02       \\
       10 & 9.484        & 5.222E-03   &  9.481       & 1.196E-02       \\
       20 & 19.492       & 2.674E-03   &  19.490      & 6.125E-03       \\
       40 & 39.496       & 1.353E-03   &  39.495      & 3.100E-03       \\
       50 & 49.497       & 1.085E-03   &  49.496      & 2.486E-03       \\
      100 & 99.498       & 5.453E-04   &  99.498      & 1.249E-03       \\
     1000 & 999.500      & 5.477E-05   &  999.500     & 1.255E-04       \\
    10000 & 9999.500     & 5.480E-06   &  9999.500    & 1.255E-05       \\
    	\hline
\end{tabular}
\caption{Critical values $(\mu/U)_{\rm c}$ and $(J/U)_{\rm c}$ for various 
        filling factors~$g$. For locating the tip of the respective Mott lobe, 
        $\mu/U$ has been varied in steps of $0.001$. Relative errors of 
        $(J/U)_{\rm c}$ are less than $1$\% in the triangular case, 
	and about $2$\% for the hexagonal lattice.}
\label{tab:J_c}
\end{table}

\begin{figure}
\centering
\includegraphics[scale=0.32,angle=-90]{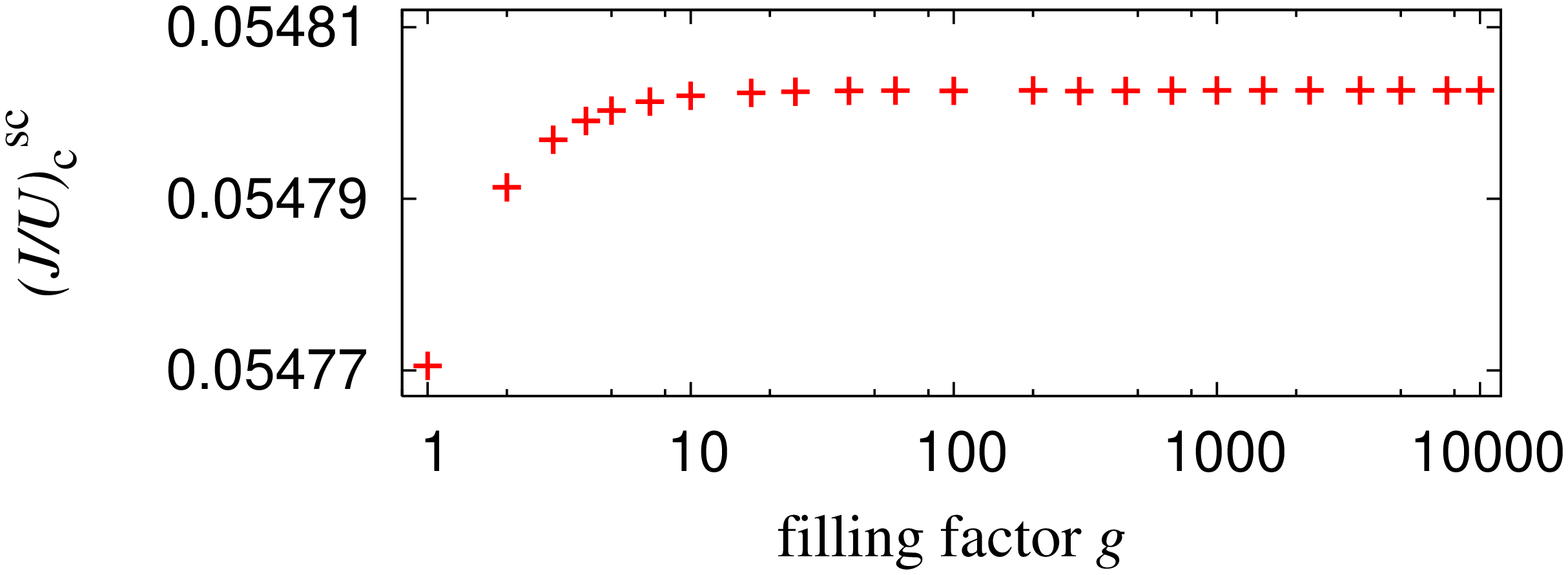}
\includegraphics[scale=0.32,angle=-90]{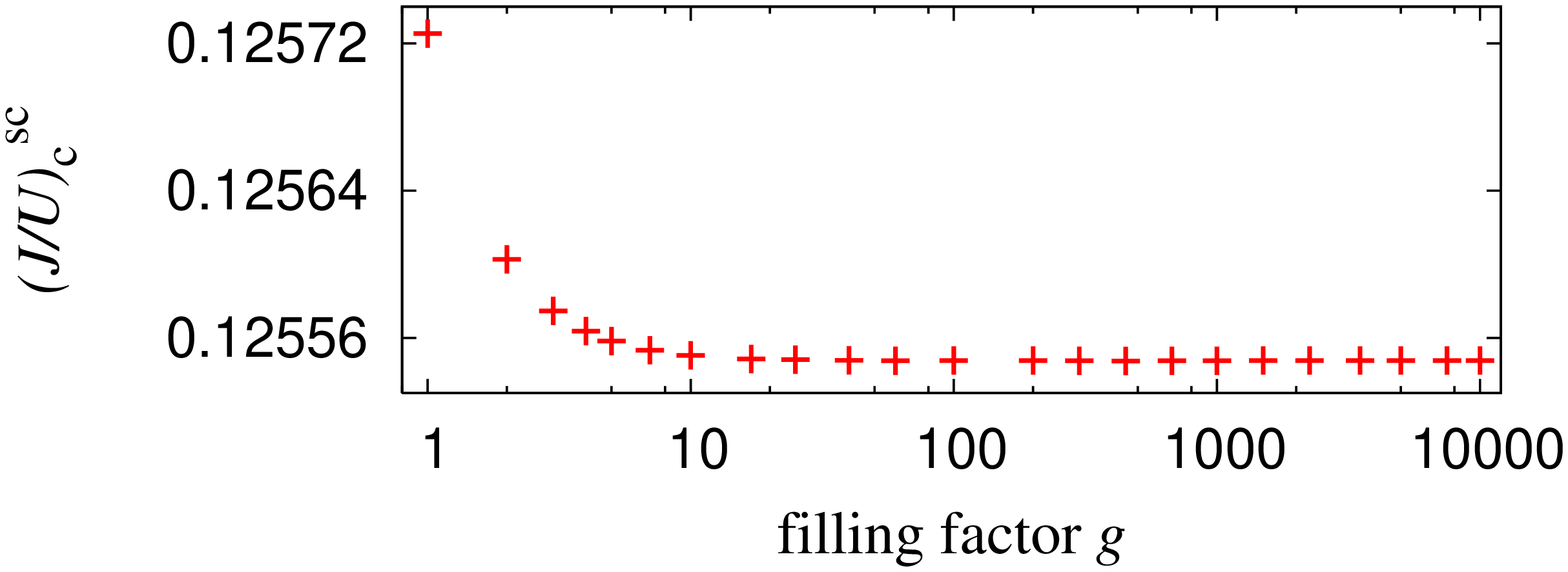}
\caption{(Color online) Scaled critical values $(J/U)_{\rm c}^{\rm sc}$ 
	according to eq.~(\ref{eq:Jcrit_scaled}) for the triangular (upper 
	plot) and for the hexagonal lattice (lower plot) vs.\ the filling 
	factor $g$. Note the very fine scale of the ordinate.}
\label{fig:Jc_scaled}
\end{figure}

\begin{figure}
\centering
\includegraphics[scale=0.32,angle=-90]{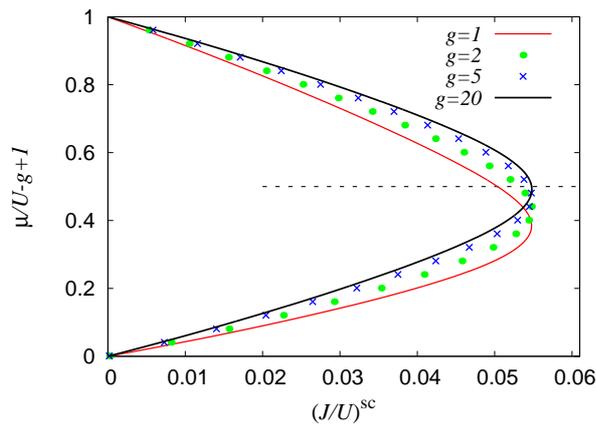}
\caption{(Color online) Scaled phase diagrams for the triangular lattice 
	with various filling factors~$g$. The scaling provided by 
	eq.~(\ref{eq:Jcrit_scaled}) stretches the different Mott lobes such 
	that their tips fall at the common value $(J/U)_{\rm c}^{\rm sc}$.
	The remaining differences are caused by the particle-hole asymmetry 
	of the Bose-Hubbard Hamiltonian. Again, the dashed horizontal line 
	$\mu/U = g - 0.5$ marks the axis of symmetry which shows up for
	sufficiently large~$g$.}
\label{fig:Scaled_phasediagrams}
\end{figure}

Figure~\ref{fig:phasediagrams} also illustrates that the critical values 
$(J/U)_{\rm c}$ decrease with increasing filling factor~$g$. As we have shown 
previously~\cite{TeichmannHinrichs2009}, in the cases of the 2D square and the 
3D cubic lattices the $g$-dependence of the exact critical values is quite well
captured by the mean-field expression~\cite{Fisher1989} for $(J/U)_{\rm c}$, 
even though the numerical agreement of the mean-field solution with the exact 
data is only moderate. Thus, the scaled critical values
\begin{equation}
	(J/U)_{\rm c}^{\rm sc} = \sqrt{g(g+1)} \left[ \frac{1}{2} +
	\sqrt{\frac{1}{4} + \frac{1}{16g(g+1)}}\, \right] (J/U)_{\rm c}
\label{eq:Jcrit_scaled}
\end{equation} 
are almost independent of~$g$. Here we demonstrate that this finding also
applies to the triangular and to the hexa\-gonal lattice by plotting in 
fig.~\ref{fig:Jc_scaled} the scaled data for both cases. As testified
by the rather fine scale of the ordinate these scaled data are practically
constant, with their residual variation amounting to only about $0.1$\%, 
which is an order of magnitude smaller than the estimated relative error 
committed in our present process-chain calculation. Finally,   
fig.~\ref{fig:Scaled_phasediagrams} shows the triangular-lattice Mott lobes
after applying the scaling~(\ref{eq:Jcrit_scaled}) not only to $(J/U)_{\rm c}$,
but to the entire phase boundaries. The scaled boundaries associated with 
different filling factors are quite similar; the remaining differences can be 
traced mainly to the particle-hole asymmetry of the Bose-Hubbard Hamiltonian.
Naturally, this asymmetry is reduced with increasing~$g$.

\section{Conclusion} 

We have presented fairly accurate phase boundaries for the homogeneous 
Bose-Hubbard model at zero temperature on both a triangular and on a 
hexagonal planar lattice, for filling factors ranging from unity to 
values so high that particle-hole symmetry is practically restored. The 
calculation has made use of the process-chain approach~\cite{Eckardt2009}, 
which already had proven its high fidelity for simple cubic 
lattices~\cite{TeichmannEtAl2009_1,TeichmannEtAl2009_2}. Our numerical 
results can serve as benchmark data for other theoretical approaches, and 
guide upcoming experiments with ultracold atoms in triangular and hexagonal 
optical lattices~\cite{BeckerEtAl2009}. Furthermore, we have shown that 
the mean-field scaling~(\ref{eq:Jcrit_scaled}) of the critical values 
$(J/U)_{\rm c}$ renders these data almost independent of the filling 
factor for both lattice types considered here. This $g$-independence of 
the data scaled in this manner thus appears to be a general feature of the 
Bose-Hubbard model, without being restricted to particular lattice geometries, 
while the lattice-specific scaled values $(J/U)_{\rm c}^{\rm sc}$ themselves 
may warrant further deliberations.

\acknowledgements 

N.T.\ wishes to thank T.P.\ Polak for stimulating discussions. Moreover,
financial support by the Deutsche Forschungsgemeinschaft (DFG) under grant
No.\ HO~1771/5 is gratefully acknowledged. Computational ressources 
have been provided by the GOLEM~I cluster of the Universit\"at Oldenburg.



\end{document}